\documentclass{desyproc}

\usepackage{graphics}
\definecolor{contribID18_vdcplotgrey}{rgb}{0.956863, 0.956863, 0.956863}

\begin{document}
\title{Drift velocity and pressure monitoring of the CMS muon drift chambers}

\author{{\slshape Lars Sonnenschein$^1$ \rm on behalf of the CMS collaboration}\\[1ex]
$^1$RWTH Aachen University, III. Physikalisches Institut A, 52056 Aachen, Germany \\[1ex]
Conference report CMS CR-2010/010, submitted to proceedings of Lepton Photon 2009
}

\contribID{18}

\acronym{LP09} 

\maketitle

\begin{abstract}
The drift velocity in drift tubes of the CMS muon chambers is a key parameter for the muon 
track reconstruction and trigger. It needs to be monitored precisely in order to detect any deviation 
from its nominal value. A change in absolute pressure, a variation of the gas admixture or a 
contamination of the chamber gas by air affect the drift velocity. 
Furthermore, the temperature and magnetic field influence its value. 
First data, taken with a dedicated Velocity Drift Chamber (VDC) 
built by RWTH Aachen IIIA are presented.
Another important parameter to be monitored is the pressure inside the muon drift tube chambers. 
The differential pressure must not exceed 
a certain value and the absolute pressure has to be kept slightly above ambient pressure to 
prevent air from entering into the muon drift tube chambers in case of a leak. 
Latest drift velocity monitoring results are discussed.
\end{abstract}

\section{Introduction}
The muon drift tube chambers of the CMS detector rely crucially on the accurate knowledge of the drift
velocity to reconstruct muon tracks as precise as possible. A monitor system for the direct 
measurement of the drift velocity is presented together with chamber gas pressure monitoring.
In the first section the CMS muon drift tube chambers are explained, followed by two sections about
their pressure and drift velocity monitoring, including latest measurements.

\section{CMS muon drift tube chambers}
The barrel of the CMS detector~\cite{cms} consists of five wheels, each instrumented with 50 muon Drift
Tube (DT) stations outside the 4~T solenoid magnet, arranged in 
four radial layers embedded in the iron return yoke of the solenoid.
Each station is 29~cm thick and has a length along the beam axis of 2.5~m, 
given by the wheel width.
Depending on the layer the station width varies between two and four meters.
The three inner muon stations consist of three and the outermost of two superlayers which in turn consist of four layers of drift cells.
The drift cells of the inner- and outermost superlayers are oriented parallel to the beam axis to measure
the projection of muon tracks perpendicular to the beam axis. The middle superlayer (present in the three inner muon stations) is dedicated to the
measurement of the track projection along the beam axis.
The drift cells are arranged with a pitch of 42~mm $\times$ 13~mm.
Field shaping strips (+1800~V) at the inner bottom and top of the drift cell are responsible for a 
very homogeneous electric field between the anode (+3600~V) and the cathode (-1200~V). 
The drift cells are operated with a gas admixture of $Ar$ (85\% vol.) and $CO_2$ (15\% vol.) at 
slight overpressure ($\sim$ 2 - 10~mbar).   
If a muon crosses a cell, ionised molecules drift to the cathode while electrons reach the anode sense wire.
The crossing position of the muon can be computed by means 
of the drift velocity assuming linear behaviour
due to a homogeneous electric field.  

\section{Pressure monitoring}
The drift tubes are operated at slight overpressure of about 2 - 10~mbar.
There are two gas manifolds at each chamber, one at the inlet and one at the outlet side.
Each manifold is equipped with two pressure sensors, one with $\pm100$~mbar and one with $\pm500$~mbar
relative pressure range, producing an analog signal of 0 - 4.5~V which is being digitised 
locally to ten bit precision, in modules called PADC's.
The pressure monitoring allows one to limit the overpressure to the range of $0 < p < 20$~mbar.
This is important to avoid contamination by ambient air and at the same time to stay well within the safety
margin. Larger leakages can be detected by input - output differences. The drift velocity measurement
in the Velocity Drift Chamber (VDC) can be corrected for the measured overpressure.   
Furthermore, each chamber is equipped with temperature sensors to allow for monitoring and corrections.

\section{Drift velocity monitoring}
\begin{figure}[b]
\unitlength 1cm
\begin{picture}(8.0, 5.5)
\put(1.5,2.7){\parbox[th]{27ex}{ 
    \caption{ \label{scetch}
      Principle and geometry of the Velocity Drift Chamber (VDC).
  }}}

\put(3.3,-7.5){\includegraphics[width=10.0cm]{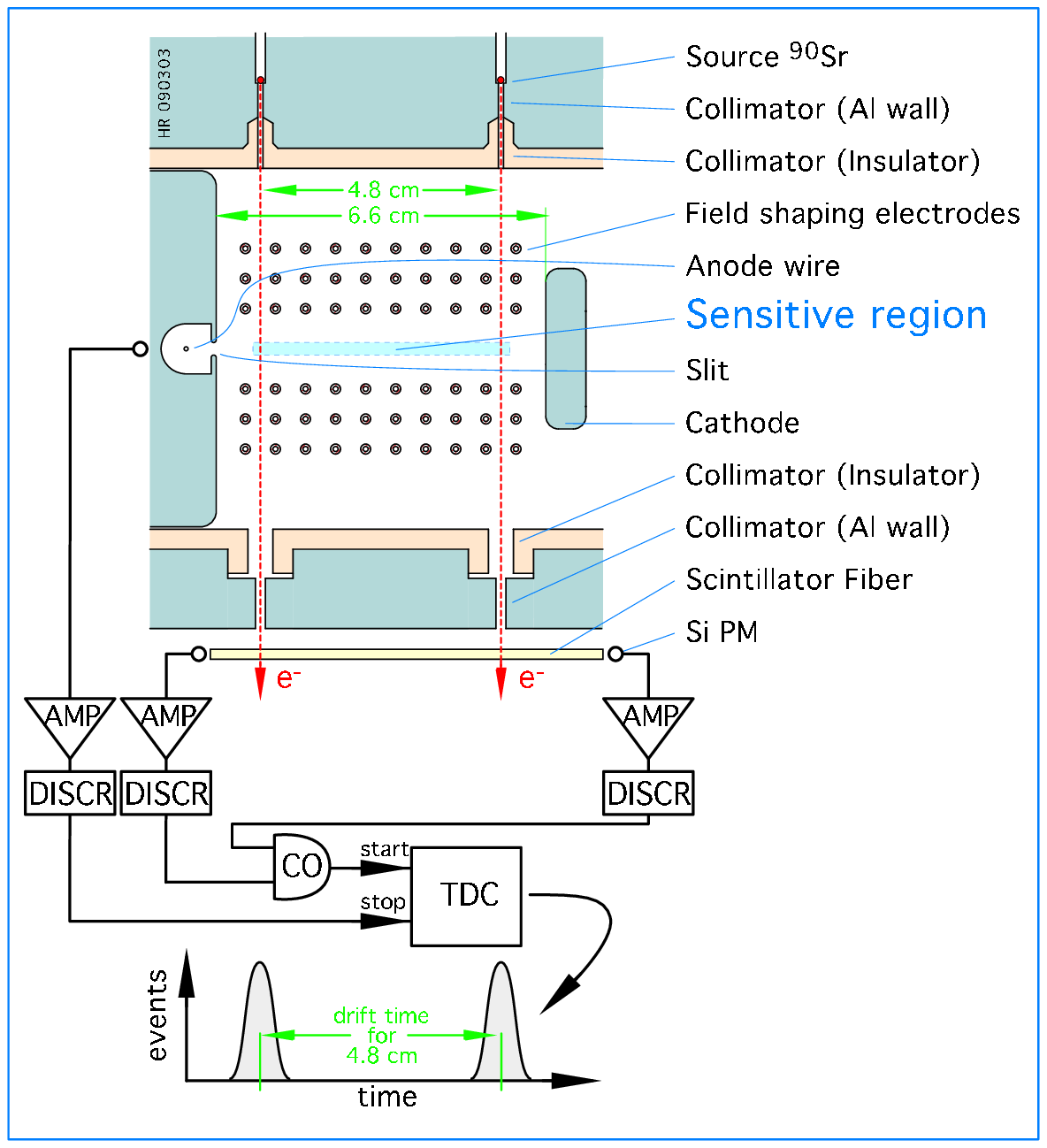}}

\end{picture}

\end{figure}

\begin{figure}[t]
\hspace*{-2.0ex}
\unitlength 1cm
\begin{picture}(8.0, 2.0)

\put(0.3, 0.05){\includegraphics[width=6.8cm]{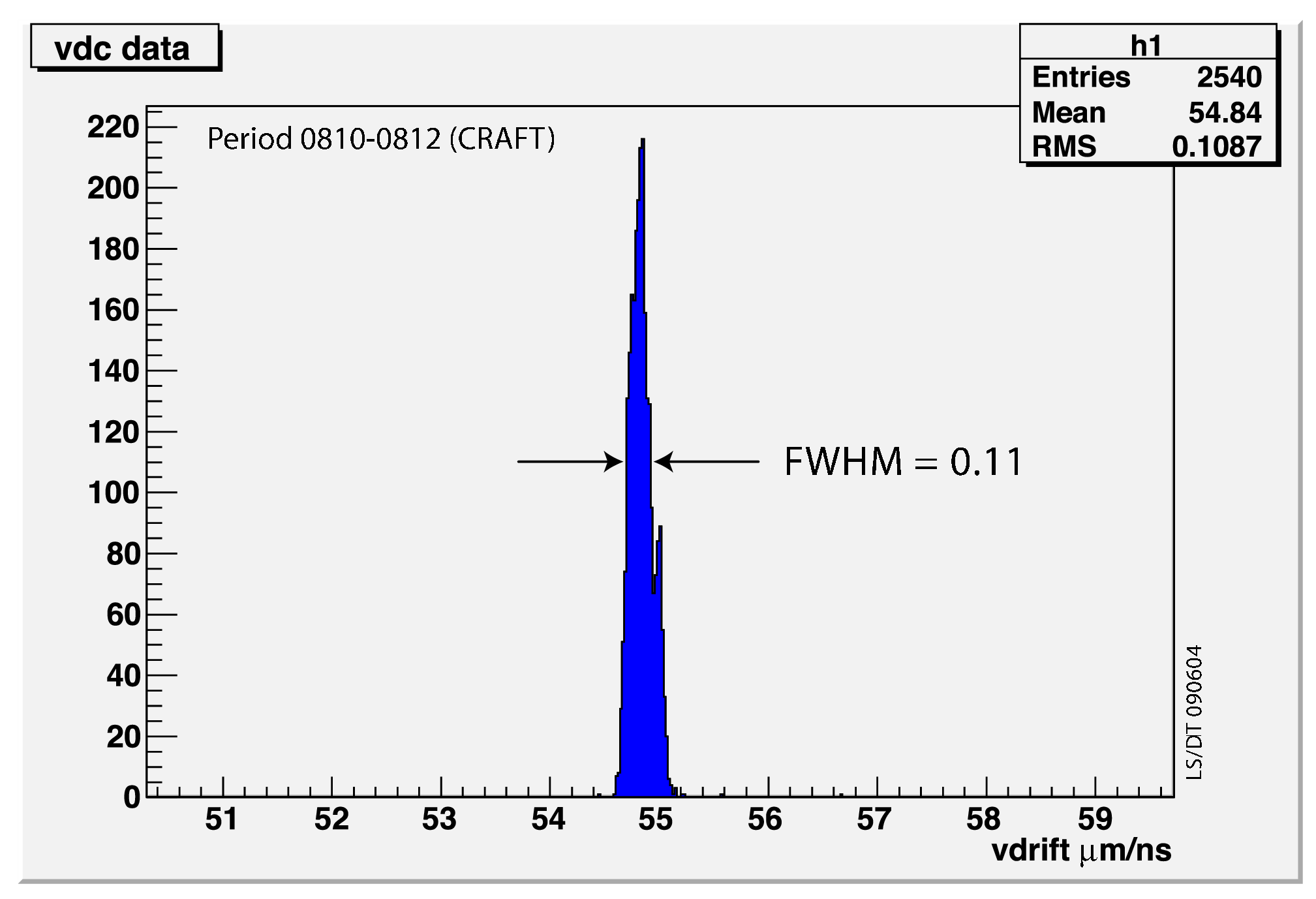} }

\put(4.3,2.27){\fcolorbox{contribID18_vdcplotgrey}{contribID18_vdcplotgrey}{ \parbox{10ex}{\vspace*{0.5ex} 
$\sigma = 0.1$ \vspace*{0.5ex}}}}
\put(6.7,1.0){\fcolorbox{contribID18_vdcplotgrey}{contribID18_vdcplotgrey}{ \parbox{9ex}{\vspace*{8ex}}}}
\put(2.82,2.27){\fcolorbox{contribID18_vdcplotgrey}{contribID18_vdcplotgrey}{ \parbox{9ex}{\vspace*{0.5ex}\phantom{aa}}}}
\put(1.5,2.8){\parbox[t]{10ex}{ VDC \\
  $V_d =$ \\ $54.8\pm 0.1~$ \\ $\mu$m/ns}}
\end{picture}
\hspace*{-3.5ex}
\includegraphics[width=6.9cm, bb=194 638 391 772, clip]{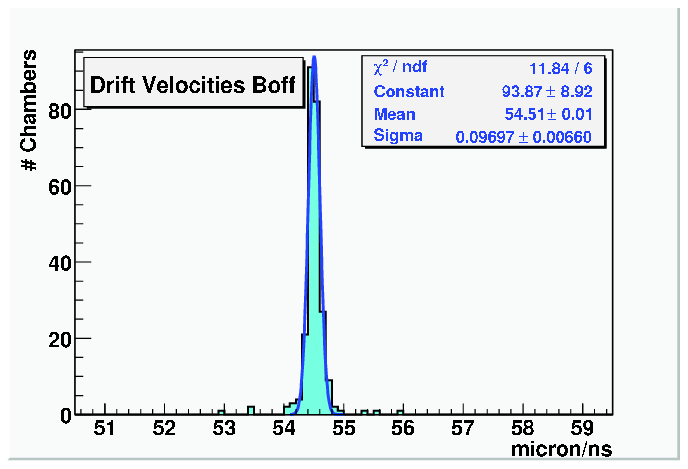}
\vspace*{-3.0ex}
\caption{ \label{vdrift}
The drift velocity measured with a VDC at CMS (left) in comparison to the drift 
velocity of the drift tube chambers (right) determined by means of measured cosmic muon tracks 
with solenoid magnetic field switched off. The plots show average values of the different muon 
stations and wheels.
}
\end{figure}

A gas sample from the DT's is piped to the VDC monitoring system 
and to commercial oxygen and humidity analysers,
to check for gas anomalies.
The outlet of each of the 250 DT chambers, as well as the supply gas arriving 
at each barrel wheel can be individually sampled.
Fig. \ref{scetch} shows the working principle of a VDC.
It consists of a small (one litre) single drift cell chamber equipped with anode (up to +1900~V), cathode 
(up to -15000~V) and field shaping electrodes.
Two electron beams cross a sensitive region between anode and cathode. The sensitive region is 
characterised by a highly uniform electric field. 
The electrons are detected by a scintillating fibre and two silicon photomultipliers (SiPM's)
in coincidence to provide a trigger. The drift times of the two beams are recorded with respect to the trigger.
The average drift time between both parallel beams, separated by 4.8~cm is being measured 
and leads to a drift velocity distribution as e.g. shown in Fig. 
\ref{vdrift}, left to 54.8~$\mu$m/ns with a width of 0.1~$\mu$m/ns by means of a single VDC
which is in operation at CMS. The measured drift velocity corresponds to two days of data taking during the
cosmic muon run period of CMS in August 2008. This can be compared to the 
indirectly determined drift velocity in the DT's
of $54.5$~$\mu$m/ns with a width of 0.1~$\mu$m/ns obtained from the cosmic muon data taking period 2008.  
The electric field of the VDC can be varied in a wide range without losing its homogeneity. The VDC measures 
the drift velocity directly for a given electric field while in the DT's the determined drift velocity is an
effective parameter depending beside the gas admixture on the inhomogeneous electric field, muon track paths, software selection and fit algorithms. Therefore the drift velocity measured with the VDC can be exploited to verify the gas admixture and the indirectly determined drift velocity of a given muon station.
Since January 2008 a single VDC unit is in operation at CMS, and it will be substituted by a larger system.

The complete VDC system consists of two racks.
The first rack contains six VDC's (one for each wheel plus a spare one), 
flow controllers, pressure controllers and sensors, 
high voltage (HV) for the VDC's, a trigger for each VDC
and a crate with dedicated VME modules.
%
The second rack contains VME logics, a readout PC (Linux server), 
NIM modules, an UPS protection power supply and a gas crate.

Determination of the systematic uncertainties of the drift velocities measured by the six VDC's 
simultaneously while branched on one single gas circuit are on the way. Furthermore 
calibration runs for HV, PM voltage, gas admixtures, impurities ($O_2, N_2$), pressure and temperature dependencies 
are 
in the process of being accomplished. \\
In summary, gas admixture anomalies and the drift velocity of the DT's in the muon stations can be verified and 
validated by means of the VDC monitoring system which measures the drift velocity directly.



\section{Bibliography}


\begin{footnotesize}




\begin{thebibliography}{99}


\bibitem{cms} CMS collaboration, 
\emph{The CMS experiment at the CERN LHC},
JINST 3 (2008) S08004.

\end{thebibliography}
%

\end{footnotesize}


\end{document}